\documentstyle[12pt]{article}
\begin{document}
\hfill{NCKU-HEP-98-02}\par
\vskip 0.5cm
\begin{center}
{\large {\bf Gauge invariance of the resummation approach to
evolution equations}}\par
\vskip 1.0cm
Hsiang-nan Li
\vskip 0.5cm
Department of Physics, National Cheng-Kung University, \par
Tainan, Taiwan, Republic of China
\end{center}
\vskip 1.0cm

%PACS numbers: 12.38.Cy, 11.10.Hi
\vskip 1.0cm
%\baselineskip=2\baselineskip

\centerline{\bf Abstract}
\vskip 0.3cm
We show that the Collins-Soper-Sterman resummation approach to the
derivation of the Dokshitzer-Gribov-Lipatov-Altarelli-Parisi equation
is gauge invariant. The special gauge-dependent parton distribution
function employed in the resummation technique is expressed as the
convolution of an infrared finite function with the standard distribution
function. By means of this convolution relation, we explain how the
technique works in summing large logarithmic corrections, and how the gauge
invariance of the special distribution function is restored.

\newpage
\centerline{\large\bf 1. Introduction}
\vskip 0.5cm

It is known that a quark distribution function for a hadron in
the minimal subtraction scheme is defined by
\begin{equation}
\phi(\xi,\mu/m)=\int\frac{dy^-}{2\pi}e^{-i\xi p^+y^-}
\langle p| {\bar q}(y^-)\frac{1}{2}\gamma^+ Pe^{i\int_0^{y^-}dz
v'\cdot A(zv')}q(0)|p\rangle\;,
\label{dep}
\end{equation}
where $\gamma^+$ is a Dirac matrix, and $|p\rangle$ denotes the incoming
hadron with the momentum $p^\mu=p^+\delta^{\mu+}$. Averages over spins
and colors are understood. The above expression, with the presence of the
path-ordered exponential $Pe^{i\int dzv'\cdot A(zv')}$,
$v^{'\mu}=\delta^{\mu-}$ being a light-like vector, is gauge invariant.
Through this exponential, $\phi$ collects the collinear
divergences of radiative correcctions to a QCD process. While the infrared
finite piece of radiative corrections is absorbed into a hard scattering
amplitude. The factorization formula for a cross section (or a structure
function) is then expressed as the convolution of these two factors.
$\phi(\xi,\mu/m)$ describes the probability that a parton carries the
fractional momentum $\xi p$ at the factorization (or renormalization)
scale $\mu$, with $m$ an infrared cutoff for the collinear divergences,
such as quark mass. The argument $\mu/m$ denotes the large logarithms
$\ln (\mu/m)$ from the collinear divergences, which will be summed by the
Dokshitzer-Gribov-Lipatov-Altarelli-Parisi (DGLAP) equation \cite{AP}.

To organize the large logarithms contained in the distribution function
using the Collins-Soper-Sterman resummation technique \cite{CS}, the vector
$v'$ is replaced by an arbitrary vector $n$, $n^2\not= 0$, leading to a
$n$-dependent distribution function,
\begin{eqnarray}
\phi^{(n)}(\xi,\nu/\mu,\mu/m)=\int\frac{dy^-}{2\pi}e^{-i\xi p^+y^-}
\langle p| {\bar q}(y^-)\frac{1}{2}\gamma^+
Pe^{i\int_0^{y^-}dz n\cdot A(zn)}q(0)|p\rangle\;.
\label{den}
\end{eqnarray}
We have made explicit the additional argument
$\nu=\sqrt{(p\cdot n)^2/|n^2|}$ of $\phi^{(n)}$, which appears as a ratio
because of the scale invariance of Eq.~(\ref{den}) in $n$ \cite{L1}.
Unfortunately, this replacement spoils the gauge invariance of 
Eq.~(\ref{dep}), since, with $n$ being
arbitrary, the end point of the path, on which the gauge field $A$ is
evaluated, does not coincide with the coordinate $y^-$ of the quark
field ${\bar q}$. It will be shown that the scale $\nu$ serves as an
ultraviolet cutoff for the loop integrals associated with the collinear
gluons, so that $\phi$ and $\phi^{(n)}$ possess the same infrared
structures but different ultraviolet structures.

We shall demonstrate that the evolution kernel of the DGLAP equation,
obtained from the resummation technique, turns out to be $n$-independent,
{\it i.e.}, gauge invariant. After deriving the equation, $n$ is brought
back to the light cone. In the $n\to v'$ limit $\nu$ diverges, and
$\ln(\nu/\mu)$ corresponds to an ultraviolet pole in dimensional
regularization. This pole is removed in a renormalization scheme, and
$\phi^{(n)}$ reduces to the gauge invariant distribution function $\phi$.
We also show that $\phi^{(n)}$ approaches $\phi$ as $\nu=\mu$, for which
the extra logarithms $\ln(\nu/\mu)$ vanish. That is, the arbitrary vector
$n$ appears only at the intermediate stage of the formalism, and as an
auxiliary tool. The gauge invariance of the resummation approach to the
evolution equation is thus guaranteed. To make our reasoning concrete, we
derive the DGLAP equation in both covariant and axial gauges.
The above discussion can be extended to the gluon distribution
function directly, whose evolution in the small momentum fraction is
governed by the Balitsky-Fadin-Kuraev-Lipatov (BFKL) equation \cite{BFKL}.

We explore the relation between the
working definition $\phi^{(n)}$ and the standard definition $\phi$: the
former is expressed as the convolution of an infrared finite
$\nu$-dependent function with the latter, because they contain the same
nonperturbative information. Hence, varying $n$ can be understood as
different ways to partition the infrared finite contribution to a cross
section between the parton distribution function and the hard scattering
amplitude. The $n$ dependence is then in fact a factorization scheme
dependence. By means of this convolution relation, we analyze in details
how the resummation for $\phi^{(n)}$ is accomplished, and how $\phi^{(n)}$
becomes $\phi$ in the limit $n\to v'$.

\vskip 1.0cm

\centerline{\large\bf 2. The DGLAP Equation}
\vskip 0.5cm

Take deep inelastic scattering of a hadron as an example. We first explain
how to factorize collinear gluons into the quark distribution function, and
justify the replacement of $v'$ by $n$ in the collinear
kinematic region \cite{L2}. It is known that the collinear region with loop
momenta of radiative gluons parallel to $p$ is important, from which large
logarithms arise \cite{L1}. In this region the partial integrand for a loop
diagram, with the radiative gluon attaching the scattered quark line, is
approximated by
\begin{equation}
\not p'\gamma_\mu\frac{\not p'+\not l}{(p'+l)^2}\approx
\not p'\frac{p'_\mu}{p'\cdot l}=
\not p'\frac{v'_\mu}{v'\cdot l}\;,
\label{e1}
\end{equation}
where the scattered quark momentum $p'$ possesses a large minus component
$p^{'-}$ in the large $x$ limit. The preceding factor $\not p'$ comes from
the final-state cut. Equation (\ref{e1}) indicates that the scattered quark
line, the collinear gluons attach, can be replaced by an eikonal line in the
direction $v'$: $1/(v'\cdot l)$ is the Feynman rule for an eikonal
propagator, and $v'_{\mu}$ for a vertex on the eikonal line, $l$ being the
momentum flowing through it. It is straightforward to show that these rules
are exactly produced by the path-ordered exponential in Eq.~(\ref{dep}).
With the eikonalization, the collinear gluons are factorized into the quark
distribution funtion $\phi$.

We may employ the further approximation,
\begin{equation}
\frac{v'_\mu}{v'\cdot l}=\frac{\delta_{-\mu}}{l^+}=
\frac{n^-\delta_{-\mu}}{n^- l^+}\approx
\frac{n^-\delta_{-\mu}+n^+\delta_{+\mu}+{\bf n}_T\delta_{T\mu}}
{n^- l^++n^+l^--{\bf n}_T\cdot {\bf l}_T}=
\frac{n_\mu}{n\cdot l}\;.
\label{e2}
\end{equation}
The smaller components $l^-$ and $l_T$ have been added to the denominator, 
and the terms proportional to the components $n^+$ and $n_T$ of an
arbitrary vector $n$, which give vanishing contributions when contracted
with a vertex in the distribution function, have been included in the
numerator. Similarly, the path-ordered exponential in $\phi^{(n)}$
generates the Feynman rules $n_\mu/(n\cdot l)$ in Eq.~(\ref{e2}). Because
$\phi^{(n)}$ and $\phi$ coincide with each other in the collinear region,
they contain the same nonperturbative information, and
their difference is infrared finite as stated in the Introduction. A
general diagram for $\phi^{(n)}$ is exhibited in Fig.~1(a), where the
exponential is represented by an eikonal line along $n$ with collinear
gluons attaching it.

In the resummation framework the DGLAP equation can be obtained by studying
the derivative $p^+d\phi^{(n)}/dp^+$ in the covariant gauge
$\partial\cdot A=0$. Due to the argument $\nu$, we have the relation
\cite{CS,L1},
\begin{eqnarray}
p^+\frac{d}{dp^+}\phi^{(n)}=-\frac{n^2}{v\cdot n}v_\alpha
\frac{d}{dn_\alpha}\phi^{(n)}\;.
\label{cph}
\end{eqnarray}
with $v^\mu=\delta^{\mu+}$ a vector along $p$. Since $p$ flows through
all the quark and gluon lines in the distribution function, while $n$
appears only in the exponential, Eq.~(\ref{cph}) simplifies the analysis
of the $p^+$ dependence of $\phi^{(n)}$. This is the motivation to introduce
the arbitrary vector $n$. The differentiation of Eq.~(\ref{e2}) with
respect to $n_\alpha$,
\begin{eqnarray}
-\frac{n^2}{v\cdot n}v_\alpha\frac{d}{dn_\alpha}\frac{n_\mu}{n\cdot l}
&=&\frac{n^2}{v\cdot n}\left(\frac{v\cdot l}{n\cdot l}n_\mu-v_\mu\right)
\frac{1}{n\cdot l}
\nonumber \\
&\equiv&\frac{{\hat n}_\mu}{n\cdot l}\;,
\label{dp}
\end{eqnarray}
gives
\begin{equation}
p^+\frac{d}{dp^+}\phi^{(n)}=2{\tilde \phi}^{(n)}\;,
\label{dph}
\end{equation}
which is described by Fig.~2(a). A summation over different attachments of
the symbol $\times$, which represents the special vertex ${\hat n}_\mu$
defined by the second expression in Eq.~(\ref{dp}), is understood. The
coefficient 2 comes from the equality of the new functions
${\tilde \phi}^{(n)}$ with the special vertex on either side of the
final-state cut.

If the loop momentum $l$ is parallel to $p$, the factor $v\cdot l$ 
vanishes, and ${\hat n}_\mu$ is proportional to $v_\mu$. When this 
$v_\mu$ is contracted with a vertex in ${\tilde \phi}^{(n)}$, where all
momenta are mainly parallel to $p$, the resultant contribution diminishes.
Therefore, the leading regions of $l$ are soft and hard, in which the
subdiagram containing the special vertex can be factorized, leading to
\begin{equation}
{\tilde \phi}^{(n)}(x,\nu/\mu,\mu/m)=
\int_x^1 d\xi [K(x/\xi,\nu/\mu)+G(x/\xi,\nu/\mu)]
\phi^{(n)}(\xi,\nu/\mu,\mu/m)\;,
\label{de1}
\end{equation}
with the functions $K$ and $G$ absorbing the soft and ultraviolet
divergences, respectively. The $O(\alpha_s)$ contributions to
$K$ from Fig.~2(b), where the eikonal approximation for the valence quark
propagator has been made, and to $G$ from Fig.~2(c), where the soft
subtraction ensures a hard loop momentum flow, are written as
\begin{eqnarray}
K&=&-ig^2{\cal C}_F\mu^\epsilon\int
\frac{d^{4-\epsilon}l}{(2\pi)^{4-\epsilon}}
\frac{{\hat n}_\mu v^\mu}{n\cdot lv\cdot l}
\left[\frac{\delta(\xi-x)}{l^2}+2\pi i\delta(l^2)
\delta\left(\xi-x-\frac{l^+}{p^+}\right)
\right]\;,\nonumber\\
& & \label{k1}\\
G&=&-ig^2{\cal C}_F\mu^\epsilon\int\frac{d^{4-\epsilon}l}
{(2\pi)^{4-\epsilon}}\frac{{\hat n}_\mu}{n\cdot l l^2}
\left[\frac{\not p+\not l}{(p+l)^2}\gamma^\mu-\frac{v^\mu}{v\cdot l}\right]
\delta(\xi-x)\;,
\label{g1}
\end{eqnarray}
${\cal C}_F=4/3$ being a color factor.

A straightforward calculation gives
\begin{eqnarray}
K&=&\frac{\alpha_s(\nu)}{\pi\xi}{\cal C}_F\left[\frac{1}{(1-x/\xi)_+}
+\ln\frac{\nu}{\mu}\delta(1-x/\xi)\right]\;,
\label{kir}\\
G&=&-\frac{\alpha_s(\nu)}{\pi\xi}{\cal C}_F\ln\frac{\xi\nu}{\mu}
\delta(1-x/\xi)\;,
\label{kgir}
\end{eqnarray}
where constants of order unity have been dropped. In the considered region
with $x\to 1$ the logarithm $\ln(\xi\nu/\mu)$ in $G$ can be replaced
by $\ln(\nu/\mu)$. Adding the above expressions, we have
\begin{eqnarray}
K(x/\xi,\nu/\mu)+G(x/\xi,\nu/\mu)
=\frac{\alpha_s(\nu)}{\pi\xi}{\cal C}_F\frac{1}{(1-x/\xi)_+}\;,
\label{skg}
\end{eqnarray}
where the argument of $\alpha_s$ has been set to the characteristic scale
$\nu$ of the evolution kernel $K+G$. It is observed that the gauge factors
$\nu$ have cancelled between $K$ and $G$.

Inserting Eq.~(\ref{skg}) into (\ref{de1}), Eq.~(\ref{dph}) becomes
\begin{equation}
p^+\frac{d}{dp^+}\phi^{(n)}(x,\nu/\mu,\mu/m)=
\frac{\alpha_s(\nu)}{\pi}{\cal C}_F
\int_x^1 \frac{d\xi}{\xi} \frac{2}{(1-x/\xi)_+}
\phi^{(n)}(\xi,\nu/\mu,\mu/m)\;.
\label{dph0}
\end{equation}
The solution to the above differential equation can be written as
\begin{eqnarray}
\phi^{(n)}(x,1,\mu/m)&=&\phi^{(n)}(x,\Lambda/\mu,\mu/m)
\nonumber \\
& &+\int_\Lambda^\mu \frac{d{\bar\mu}}{\bar\mu}
\frac{\alpha_s({\bar\mu})}{\pi}{\cal C}_F
\int_x^1 \frac{d\xi}{\xi} \frac{2}{(1-x/\xi)_+}
\phi^{(n)}(\xi,{\bar\mu}/\mu,\mu/m)\;,
\nonumber\\
& &
\label{sp1}
\end{eqnarray}
with $\Lambda$ an arbitrary scale. We shall explain in Sect. 3. that
$\phi^{(n)}$ coincides with $\phi$ as $\nu=\mu$, {\it i.e.}, as the
logarithms $\ln(\nu/\mu)$ vanish:
\begin{eqnarray}
\phi^{(n)}(x,1,\mu/m)=\phi(x,\mu/m)\;.
\label{pnp}
\end{eqnarray}
Differentiating Eq.~(\ref{sp1}) with respect to $\mu$, and employing 
the RG equation shown in Sect. 3,
\begin{equation}
\mu \frac{d}{d\mu}\phi^{(n)}(x(\xi),\Lambda({\bar \mu})/\mu,\mu/m)
=-2\lambda_q\phi^{(n)}(x(\xi),\Lambda({\bar \mu})/\mu,\mu/m)\;,
\label{rgd}
\end{equation}
with $\lambda_q=-\alpha_s/\pi$ the quark anomalous dimension, we obtain
\begin{equation}
\mu\frac{d}{d\mu}\phi(x,\mu/m)=
\frac{\alpha_s(\mu)}{\pi}{\cal C}_F\int_x^1 \frac{d\xi}{\xi} 
\frac{2}{(1-x/\xi)_+}\phi(\xi,\mu/m)
-2\lambda_q(\mu)\phi(x,\mu/m)\;.
\label{sp3}
\end{equation}
At this step, the $n$ dependence disappears completely.
The above formula is then identified as the DGLAP equation,
\begin{eqnarray}
\mu\frac{d}{d\mu}\phi(x,\mu/m)=\frac{\alpha_s(\mu)}{\pi}
\int_x^1 \frac{d\xi}{\xi} P(x/\xi)\phi(\xi,\mu/m)\;,
\label{sp2}
\end{eqnarray}
where the evolution kernel $P$ is given by
\begin{eqnarray}
P(x)={\cal C}_F\left[\frac{2}{(1-x)_+}+\frac{3}{2}\delta(1-x)\right]\;.
\end{eqnarray}

Next we derive the DGLAP equation in the axial gauge \cite{L2,L3}, in
which the relevant definitions for the quark distribution function are
\begin{eqnarray}
\phi^{(n)}(\xi,\mu/m)&=&\int\frac{dy^-}{2\pi}e^{-i\xi p^+y^-}
\langle p| {\bar q}(y^-)\frac{1}{2}
\gamma^+ q(0)|p\rangle|_{v'\cdot A=0}\;,
\label{dep1}\\
\phi^{(n)}(\xi,\nu/\mu,\mu/m)&=&\int\frac{dy^-}{2\pi}e^{-i\xi p^+y^-}
\langle p| {\bar q}(y^-)\frac{1}{2}
\gamma^+ q(0)|p\rangle|_{n\cdot A=0}\;,
\label{den1}
\end{eqnarray}
as described by Fig.~1(b). The path-ordered exponential is equal to the
identity in this gauge, and eikonal lines collecting the collinear gluons
are absent. Similarly, we work on $\phi^{(n)}$ when performing the
resummation. The $n$ dependence goes into the gluon propagator,
$(-i/l^2)N^{\mu\nu}(l)$, with
\begin{equation}
N^{\mu\nu}=g^{\mu\nu}-\frac{n^\mu l^\nu+n^\nu l^\mu}
{n\cdot l}+n^2\frac{l^\mu l^\nu}{(n\cdot l)^2}\;.
\label{gp}
\end{equation}
Because of the scale invariance of $\phi^{(n)}$ in $n$ as indicated by
Eq.~(\ref{gp}), $\phi^{(n)}$ depends on the scale $\nu$, and thus
Eq.~(\ref{cph}) holds. The operator $d/dn_\alpha$, now applying to the
gluon propagator, gives
\begin{eqnarray}
-\frac{n^2}{v\cdot n}v_\alpha\frac{d}{dn_\alpha}N^{\mu\nu}
&=&\frac{n^2v_\alpha}{v\cdot nn\cdot l}
(l^\mu N^{\alpha\nu}+l^\nu N^{\mu\alpha})\;,
\nonumber \\
&\equiv&{\hat v}_\alpha(l^\mu N^{\alpha\nu}+l^\nu N^{\mu\alpha})\;.
\label{dgp}
\end{eqnarray}
The loop momentum $l^\mu$ ($l^\nu$) carried by the differentiated gluon 
contracts with the vertex the gluon attaches, which is then replaced by
the special vertex ${\hat v}_\alpha$ defined by the second
line of Eq.~(\ref{dgp}).

The contraction of $l^\nu$ hints the application of the Ward
identities \cite{L2,L3}, such as
\begin{equation}
\frac{i(\not k+\not l)}{(k+l)^2}(-i\not l)\frac{i\not k}{k^2}
=\frac{i\not k}{k^2}-\frac{i(\not k+\not l)}{(k+l)^2}\;,
\end{equation}
for the quark-gluon vertex, and
\begin{equation}
l^\nu\frac{-iN^{\alpha\mu}(k+l)}{(k+l)^2}\Gamma_{\mu\nu\lambda}
\frac{-iN^{\lambda\gamma}(k)}{k^2}
=-i\left[\frac{-iN^{\alpha\gamma}(k)}{k^2}-
\frac{-iN^{\alpha\gamma}(k+l)}{(k+l)^2}\right]\;,
\label{ward}
\end{equation}
for the triple-gluon vertex $\Gamma_{\mu\nu\lambda}$. Similar identities
for other types of vertices can be derived easily.
Summing all the diagrams with different differentiated gluons, those 
embedding the special vertices cancel by pairs, leaving the one with
the special vertex moving to the outer end of the quark line \cite{CS}.
We obtain the same formula in Eq.~(\ref{dph}), described by Fig.~3(a),
where the new function ${\tilde \phi}^{(n)}$ contains one special vertex
represented by a square. The coefficient 2 comes from the equality of the
new functions with the special vertex on either of the two quark 
lines.

The important regions of the loop momentum $l$ flowing through the
special vertex are also soft and hard, since the vector $n$ does not lie on
the light cone, and the collinear enhancements are suppressed. Similarly,
in the above leading regions ${\tilde \phi}^{(n)}$ can be factorized into
the convolution of the subdiagram containing the special vertex with the
original distribution function $\phi^{(n)}$, leading to Eq.~(\ref{de1}).
In this case the functions $K$ and $G$, corresponding to Figs.~3(b) and
3(c), are written as
\begin{eqnarray}
K&=&ig^2{\cal C}_F\mu^\epsilon\int
\frac{d^{4-\epsilon}l}{(2\pi)^{4-\epsilon}}
\frac{{\hat v}_\mu v_\nu}{v\cdot l}
\left[\frac{\delta(\xi-x)}{l^2}+2\pi i\delta(l^2)
\delta\left(\xi-x-\frac{l^+}{p^+}\right)\right]N^{\mu\nu},
\nonumber \\
& &
\label{kj}\\
G&=&-ig^2{\cal C}_F\mu^\epsilon\int\frac{d^{4-\epsilon}l}
{(2\pi)^{4-\epsilon}}{\hat v}_\mu
\left[\frac{\xi\not p-\not l}{(\xi p-l)^2}\gamma_\nu
+\frac{v_\nu}{v\cdot l}\right]\frac{N^{\mu\nu}}{l^2}\delta(\xi-x)\;,
\label{gph}
\end{eqnarray}
It is easy to show that Eqs.~(\ref{kj}) and (\ref{gph}) reduce to
Eqs.~(\ref{kir}) and (\ref{kgir}), respectively. Then following the steps
from Eq.~(\ref{skg}) to Eq.~(\ref{sp2}), we derive the same DGLAP equation
in the axial gauge. This is natural, because the evolution of a parton
distribution function is measurable, and should be independent of the
gauge we adopt.

\vskip 1.0cm

\centerline{\large\bf 3. Relation of $\phi^{(n)}$ and $\phi$}
\vskip 0.5cm

In this section we explore the relation between the working definition
$\phi^{(n)}$ of the quark distribution function employed in the resummation
technique and the standard definition $\phi$. As explained in Sect. 2,
$\phi^{(n)}$ and $\phi$ contain the same collinear divergences from 
gluon momenta parallel to the hadron momentem, and thus the same
nonperturbative information. Take the $O(\alpha_s)$ correction with one
end of a real gluon attaching the quark line and the other end attaching
the eikonal line  as an example. The loop integral associated with
$\phi$ is written as
\begin{eqnarray}
I&=&g^2C_F\mu^\epsilon\int\frac{d^{4-\epsilon}l}{(2\pi)^{4-\epsilon}}
\frac{(x\not p+\not l){\not v}'}{(xp+l)^2 v'\cdot l}
2\pi\delta(l^2)\delta\left(1-x-\frac{l^+}{p^+}\right)\;,
\nonumber\\
&=&\frac{\alpha_s}{\pi}\frac{\mu^\epsilon}{(2\pi)^{1-\epsilon}}
\frac{x}{2(1-x)}\int d^{2-\epsilon}l_T \frac{1}{l_T^2+m^2}\;.
\label{i}
\end{eqnarray}
where $m$ is the infrared cutoff mentioned before, and $\mu$ is the
renormalization scale. The corresponding integral $I^{(n)}$ associated with
$\phi^{(n)}$ is obtained by replacing $v'$ in Eq.~(\ref{i}) by $n$, which
is given by,
\begin{eqnarray}
I^{(n)}=\frac{\alpha_s}{2\pi^2}
\int d^2l_T \frac{2x(1-x)\nu^2}{(l_T^2+m^2)[l_T^2+4(1-x)^2\nu^2]}\;.
\label{in}
\end{eqnarray}
The collinear divergences come from the region with small $l_T$,
in which $I^{(n)}$ reduces to $I$ as indicated by Eqs.~(\ref{i}) and
(\ref{in}). However, $I$ is ultraviolet divergent, while $I^{(n)}$ is not,
because of the extra denominator $l_T^2+4(1-x)^2\nu^2$. Therefore,
$\phi^{(n)}$ and $\phi$ possess the same collinear
structure but different ultraviolet structure.

Performing the integrations over $l_T$, we obtain
\begin{eqnarray}
I&=&\frac{\alpha_s}{2\pi}\frac{x}{1-x}
\left(\frac{1}{2\epsilon}+\ln\frac{\mu}{m}\right)\;,
\label{i1}\\
I^{(n)}&=&\frac{\alpha_s}{2\pi}\frac{x}{1-x}
\ln\frac{\nu}{m}\;,
\label{in1}
\end{eqnarray}
where constants of order unity have been dropped. It is found that $I$
contains a logarithm $\ln(\mu/m)$, and $I^{(n)}$ contains $\ln(\nu/m)$
with the same coefficient. Corrections in $\phi^{(n)}$ and $\phi$ without
gluons attaching the eikonal line, such as the self-energy corrections,
give the same logarithms $\ln(\mu/m)$. By splitting the logarithm in
$I^{(n)}$ into
\begin{equation}
\ln(\nu/m)=\ln(\nu/\mu)+\ln(\mu/m)\;,
\end{equation}
it is easy to understand why $\phi^{(n)}$ depends on the argument
$\nu/\mu$ due to the replacement of $v'$ by $n$. The resummation technique,
summing the first logarithm in the above expression to all orders, leads to
Eq.~(\ref{dph0}) in the same form as the (partial) DGLAP equation, that
sums the second logarithm, since the two terms possess the same
coefficients. The choice $\nu=\mu$, diminishing the first term, renders
$\phi^{(n)}$ coincide with $\phi$ as in Eq.~(\ref{pnp}), because the second
term, combined with those from other corrections without gluons attaching
the eikonal line, gives the complete logarithms $\ln(\mu/m)$ in $\phi$.
It is also easy to realize why the RG method, applying to
$\phi^{(n)}(x(\xi),\Lambda({\bar \mu})/\mu,\mu/m)$,
sums only $\ln(\mu/m)$ from other corrections as in Eq.~(\ref{rgd}),
since the corrections with gluons attaching the eikonal line, proportional
to $\ln(\nu/m)$, are $\mu$-independent. At last, in the limit $n\to v'$
the cutoff $\nu$ diverges, and $\ln(\nu/\mu)$ corresponds to the
ultraviolet pole in $I$. This pole is removed by renormalization,
and $\phi^{(n)}$ approaches $\phi$:
\begin{equation}
\lim_{n\to v'}\phi^{(n)}(x,\nu/\mu,\mu/m)=\phi(x,\mu/m)\;.
\label{lph}
\end{equation}

The infrared finite difference between $\phi^{(n)}$ and $\phi$, indicated
by $I^{(n)}-I\propto \ln(\nu/\mu)$ from Eqs.~(\ref{i1}) and (\ref{in1}),
can be computed perturbatively. In terms of Feynman diagrams in the
covariant gauge, it is the difference between the diagrams with
radiative gluons attaching the eikonal line along $n$ and the diagrams
with gluons attaching the eikonal line along $v'$. The above argument
leads to the convolution formula,
\begin{equation}
\phi^{(n)}(x,\nu/\mu,\mu/m)=\int_x^1\frac{d\xi}{\xi}
D(x/\xi,\nu/\mu)\phi(\xi,\mu/m)\;,
\label{nv}
\end{equation}
where the infrared finite piece $D$, independent of the infrared regulator
$m$, denotes the difference mentioned above. The graphic defintion of
$D$ is given in Fig.~4, where the denominator, with the eikonal line along
$v'$, subtracts the collinear divergences involved in the numerator with
the eikonal line along $n$. All the diagrams without gluons attaching
the eikonal lines cancel between the numerator and the denominator. The
bubbles contain infinite many gluon exchanges, and the external lines
represent quarks. Because of $D$, $\phi^{(n)}$ absorbes additional finite
contributions compared to $\phi$. Hence, they can be regarded as the
distribution functions defined in different factorization schemes. Varying
the vector $n$ then means varying the factorization scheme.

Based on Eq.~(\ref{nv}), the differentiation $d/dp^+$ or $d/dn$ in fact
applies to $D$:
\begin{equation}
p^+\frac{d}{dp^+}\phi^{(n)}(x,\nu/\mu,\mu/m)=
\int_x^1\frac{d\xi}{\xi}p^+\frac{d}{dp^+}D(x/\xi,\nu/\mu)
\phi(\xi,\mu/m)\;.
\label{nvd}
\end{equation}
The similar resummation procedures give
\begin{equation}
p^+\frac{d}{dp^+}\phi^{(n)}(x,\nu/\mu,\mu/m)=
\int_x^1\frac{d\xi}{\xi}\int_{x/\xi}^1\frac{dz}{z}
\frac{2}{[1-x/(\xi z)]_+}
D(z,\nu/\mu)\phi(\xi,\mu/m)\;.
\label{nvr}
\end{equation}
Using the variable change $z=y/\xi$, and interchanging the integrations
over $\xi$ and over $y$, the above equation is reexpressed as
\begin{equation}
p^+\frac{d}{dp^+}\phi^{(n)}(x,\nu/\mu,\mu/m)=
\int_x^1\frac{dy}{y}\int_{y}^1\frac{d\xi}{\xi}
\frac{2}{(1-x/y)_+}D(y/\xi,\nu/\mu)\phi(\xi,\mu/m)\;.
\label{nvr1}
\end{equation}
Employing the definition in Eq.~(\ref{nvd}), we arrive at Eq.~(\ref{dph0}).
The remaining steps of the resummation and Eq.~(\ref{lph}) in the
$n\to v'$ limit then lead to the DGLAP equation. The above formulas
elucidate the statement that the vector $n$ is an auxiliary tool, and
appears only at the intermediate stage of the derivation.
Hence, the gauge invariance of the resummation technique is guaranteed.

The discussion presented here can be generalized to the resummation-based 
derivation of the BFKL equation for the gluon distribution function 
$F(x,k_T)$ easily, which describes the probability that
a gluon from the hadron carries the longitudinal momemtum $xp^+$ and the
transverse momentum $k_T$. However, the relation between the working
definition $F^{(n)}$ and the standard definition $F$ is a
$k_T$-factorization formula \cite{J}, instead of a collinear factorization
formula in Eq.~(\ref{nv}). The details will be published elsewhere.

\vskip 1.0cm

\centerline{\large\bf 4. Conclusion}
\vskip 0.5cm

In this paper we have demonstrated how to derive the DGLAP equation using
the Collins-Soper-Sterman resummation technique, and that the results
are independent of the gauge we employed. Though we started with
the $n$-dependent parton distribution function, the evolution kernel turns
out to be gauge invariant. We have also explored
the relation between the $n$-dependent and standard distribution
functions. From this relation, we explained why the resummation technique
is a successful method to the summation of large logarithms,
and showed that the $n$-dependent definition reduces to the standard one
by either taking the $n\to v'$ limit or setting the gauge factor
$\nu$ to the renormalization scale $\mu$. This work provides a solid
ground for our previous studies on the resummation approach
to evolution equations \cite{L2,L3}.

\vskip 0.5cm
This work was supported by the National Science Council of R.O.C. under the 
Grant No. NSC-87-2112-M-006-018.

\newpage
\centerline{\large \bf Figure Captions}
\vskip 0.5cm

\noindent
{\bf FIG. 1.} Definition of the quark distribution function in (a) the
covariant gauge and in (b) the axial gauge.
\vskip 0.5cm

\noindent
{\bf FIG. 2.} (a) The derivative $p^+d\phi^{(n)}/dp^+$ in the covariant
gauge. (b) The $O(\alpha_s)$ function $K$. (c) The $O(\alpha_s)$ function
$G$.
\vskip 0.5cm

\noindent
{\bf FIG. 3.} (a) The derivative $p^+d\phi^{(n)}/dp^+$ in the axial
gauge. (b) The $O(\alpha_s)$ function $K$. (c) The $O(\alpha_s)$ function
$G$.
\vskip 0.5cm

\noindent
{\bf FIG. 4.} The graphic definition of the function $D$.

\end{document}